\documentclass[conference]{IEEEtran}
\IEEEoverridecommandlockouts

\usepackage{cite}
\usepackage{xcolor}
\usepackage{amsmath,amssymb,amsfonts}
\usepackage{algorithmic}
\usepackage{graphicx}
\usepackage{textcomp}
\usepackage[hidelinks]{hyperref}
\usepackage{textcomp}
\usepackage[gen]{eurosym}
\usepackage{multirow}
\usepackage{booktabs}
\usepackage{makecell}
\usepackage{soul}
\def\BibTeX{{\rm B\kern-.05em{\sc i\kern-.025em b}\kern-.08em
  T\kern-.1667em\lower.7ex\hbox{E}\kern-.125emX}}

\begin{document}

\title{Robustness of Local Energy Markets to Cyberattacks: Case Study of False Data Injection}

\author{
\IEEEauthorblockN{
Mehran Moradi\IEEEauthorrefmark{1},
Reza Zamani\IEEEauthorrefmark{2},
Phil Aupke\IEEEauthorrefmark{3},
Andreas Theocharis\IEEEauthorrefmark{1},
Andreas Kassler\IEEEauthorrefmark{3}\IEEEauthorrefmark{4}
}
\IEEEauthorblockA{\IEEEauthorrefmark{1}Engineering and Physics Department, Karlstad University, Karlstad, Sweden\\
E-mails: mehran.moradi@kau.se, andreas.theocharis@kau.se}
\IEEEauthorblockA{\IEEEauthorrefmark{2}Faculty of Electrical and Computer Engineering, Tarbiat Modares University, Tehran, Iran\\
E-mail: r\_zamani@modares.ac.ir}
\IEEEauthorblockA{\IEEEauthorrefmark{3}Department of Mathematics and Computer Science, Karlstad University, Karlstad, Sweden\\
E-mails: phil.aupke@kau.se, andreas.kassler@kau.se}
\IEEEauthorblockA{\IEEEauthorrefmark{4}Faculty of Computer Science, Deggendorf Institute of Technology, Deggendorf, Germany\\
E-mail: andreas.kassler@th-deg.de}
\thanks{This work was funded partially by the Swedish Energy Agency, project Solar Electricity Research Centre, Sweden (SOLVE), grant number 52693-1, and the Interreg SV-NO GränsEnergi project, grant number 20369648. Additional funding was provided by the Bavarian Government through the HighTech-Agenda (HTA).}
}
\author{
\IEEEauthorblockN{
Mehran Moradi$^{a,1}$,
Reza Zamani$^{b,2}$,
Phil Aupke$^{c,3}$,
Andreas Theocharis$^{a,4}$,
Andreas Kassler$^{c,d,5}$
}

\IEEEauthorblockA{$^{a}$ Engineering and Physics Department, Karlstad University, Karlstad, Sweden}
\IEEEauthorblockA{$^{b}$ Faculty of Electrical and Computer Engineering, Tarbiat Modares University, Tehran, Iran}
\IEEEauthorblockA{$^{c}$ Department of Mathematics and Computer Science, Karlstad University, Karlstad, Sweden}
\IEEEauthorblockA{$^{d}$ Faculty of Computer Science, Deggendorf Institute of Technology, Deggendorf, Germany}

\IEEEauthorblockA{
\parbox{0.98\textwidth}{\centering\footnotesize
Emails: $^{1}$mehran.moradi@kau.se, $^{2}$r\_zamani@modares.ac.ir, $^{3}$phil.aupke@kau.se, $^{4}$andreas.theocharis@kau.se, $^{5}$andreas.kassler@\{kau.se,th-deg.de\}
}
}

\thanks{This work was funded partially by the Swedish Energy Agency, project Solar Electricity Research Centre, Sweden (SOLVE), grant number 52693-1, and the Interreg SV-NO GränsEnergi project, grant number 20369648. Additional funding was provided by the Bavarian Government through the HighTech-Agenda (HTA).}
}
\maketitle
\begin{abstract}
Market clearing in community-based local energy markets relies
on power demand and PV forecasts, making it vulnerable to
coordinated false data injection (FDI) attacks. This paper
proposes a bilevel optimization framework to identify worst-case
bus-level FDI against interconnected multi-community
electricity markets within a distribution network. The upper
level attacker maximizes physical impact, measured by
cumulative voltage deviation, while accounting for
detectability. The lower level re-clears the interconnected
multi-community markets subject to operational and network
constraints. The bilevel model is reformulated as a single-level
mixed-integer program through Karush--Kuhn--Tucker
conditions and solved using the epsilon-constraint method to
characterize the trade-off between physical impact and
detectability. Case studies on the IEEE 33-bus system with
three communities show that even bounded, system-level
zero-sum attacks can reshape local trading, reduce
voltage-security margins, and produce asymmetric
community-level market outcomes. The results further show
that vulnerability depends strongly on the spatiotemporal
placement of falsified data rather than on uniform spreading
across buses and time periods.
\end{abstract}

\begin{IEEEkeywords}
Transactive energy, local energy market, cyber-security, false data injection attack.
\end{IEEEkeywords}

\section{Introduction}\label{sec:intro}

The increasing deployment of sensors, communication
infrastructure, intelligent agents, and distributed energy
resources (DERs) has transformed modern power distribution
systems into cyber-physical environments with active
prosumer participation. This shift has intensified interest
in transactive energy systems (TESs) and local energy markets
(LEMs), including peer-to-peer (P2P) trading and
community-based clearing, as mechanisms for coordinating
distributed participants and balancing local supply and
demand~\cite{book,myp2p,mmrr,rezap2p}. Yet the
reliance of these markets on continuous data exchange and
automated decision-making has introduced important
cybersecurity vulnerabilities. Among these, false data
injection (FDI) is particularly concerning because manipulated
information can distort market outcomes while remaining
difficult to detect~\cite{nist,ref1,ref7,ref15}.

Prior studies on FDI in TESs and LEMs can be broadly
grouped into two main streams. The first examines
scenario-based attacks in distribution-level settings,
including price and demand signal manipulation in TESs and
forecast and flexibility tampering in TEM-based
microgrids~\cite{ref11,ref3}. Such studies offer useful
insights into system behavior under fixed attack patterns,
but generally do not characterize worst-case adversarial
strategies. The second stream focuses on local P2P and
community trading and develops attacks that exploit market
structure. Islam \textit{et al.}~\cite{isl} formulate an
optimized, stealth-constrained FDI that maximizes attacker
benefit in a residential microgrid while preserving
supply--demand balance. Mohammadi \textit{et al.}~\cite{ref5}
show that coordinated low-magnitude demand falsification
across many prosumers can shift prices and weaken prosumer
incentives. Thota \textit{et al.}~\cite{ref6} propose a
stealth-constrained zero-sum FDI for supply-demand-ratio-based
residential P2P trading. More broadly, bilevel attacker--operator
formulations are well established in cyber-physical
power-system vulnerability analysis~\cite{ref1}.

Unlike prior works that focus on scenario-based attacks in distribution-level settings or optimized attacks in single-community P2P trading, the present study explicitly models worst-case, stealth-constrained FDI on bus-level demand and PV forecasts in interconnected multi-community LEMs embedded in a distribution network. The injected falsified demand and PV signals are bounded by forecast uncertainty margins and preserve system-wide totals at each time step, thereby maintaining aggregate stealth while redistributing perturbations across buses. The key novelty lies in demonstrating how such localized falsifications propagate through interconnected community market re-clearing and network operation, leading to voltage security degradation and asymmetric community-level outcomes.

Specifically, the paper develops a bilevel attacker--operator
model for distribution networks with multiple interconnected
energy communities. The upper level determines bounded
perturbations to demand and PV forecasts that maximize
cumulative voltage deviation while accounting for
detectability. The lower level re-clears the multi-community
electricity market through a linearized distribution optimal
power flow model subject to DER limits, nodal power
balance, and voltage constraints.

The main contributions of this paper are threefold:
1) a bilevel attacker--operator framework for coordinated
FDI targeting bus-level demand and PV forecasts in
interconnected multi-community electricity markets under
linearized distribution-network constraints;
2) evidence that even system-level zero-sum FDI attacks,
while evading aggregate consistency checks, can produce
asymmetric community-level market outcomes and uneven
impacts on voltage-security margins; and
3) a demonstration that attack severity is governed by where and when falsified data are concentrated, not by their aggregate magnitude alone, motivating detection approaches that focus on concentrated anomaly patterns rather than only aggregate consistency checks.

The remainder of this paper is organized as follows.
Section~\ref{sec:model} presents the proposed model,
Section~\ref{sec:results} discusses the results, and
Section~\ref{sec:conclusion} concludes the paper.
\section{Proposed Model}\label{sec:model}
The interconnected market comprises bus-level players
$i\in\mathcal{I}$ grouped into communities $c\in\mathcal{C}$,
with $\mathcal{I}_c\subseteq\mathcal{I}$.
Players may have demand, PV, and dispatchable DGs and trade within
communities and with the grid, while communities exchange energy over
$t\in\mathcal{T}$.
To quantify the vulnerability of interconnected multi-community electricity markets to coordinated FDI, the problem is formulated as a bilevel attacker–operator model. The attacker is assumed to have full system and market knowledge, providing a conservative worst-case vulnerability benchmark. The upper level determines bounded perturbations to demand and PV forecasts, whereas the lower level re-clears the interconnected multi-community markets under operational and distribution-network constraints using the manipulated data.
\begin{equation}\label{eq:upper_obj}
\left\{
\begin{array}{ll}
\max & \underbrace{\Phi\!\left(\mathbf{V}(\tilde d,\tilde p^{\mathrm{PV}})\right)}_{\text{voltage deviation}} \\[0.4em]
\min & \underbrace{\displaystyle \sum_{t\in\mathcal{T}}\sum_{i\in\mathcal{I}}
\left(\left|\Delta d_{i,t}\right| + \left|\Delta p^{\mathrm{PV}}_{i,t}\right|\right)}_{\text{detectability index}}
\end{array}
\right.
\end{equation}
\subsection{Attacker's Strategy}\label{sect:upper}
The attacker's FDI strategy, shown in Eq. \eqref{eq:upper_obj}, is formulated as a multi-objective problem that maximizes physical impact while minimizing the detectability index. Physical impact is measured by the cumulative voltage deviation $\Phi(\mathbf{V}(\tilde d,\tilde p^{\mathrm{PV}})) \triangleq \sum_{t\in\mathcal{T}}\sum_{i\in\mathcal{I}} |V_{i,t}-V^{\mathrm{ref}}|$, where $\mathcal{I}$ and $\mathcal{T}$ denote the sets of network buses and periods, and $V^{\mathrm{ref}}$ denotes the nominal per-unit voltage, set to 1.0 p.u. Here, $\mathbf{V}(\tilde d,\tilde p^{\mathrm{PV}})$ is the voltage profile obtained after lower-level market re-clearing under the manipulated demand and PV inputs $\tilde d$ and $\tilde p^{\mathrm{PV}}$. The detectability index is quantified as the aggregate absolute magnitude of the perturbations injected into the forecasted demand and PV inputs.

The adversary injects additive perturbations $\Delta d_{i,t}$ and $\Delta p^{\mathrm{PV}}_{i,t}$ into the original demand and PV generation forecasts, denoted $d_{i,t}$ and $p^{\mathrm{PV}}_{i,t}$, respectively, so that the market operator receives the manipulated signals:
\begingroup
\setlength{\abovedisplayskip}{3pt}
\setlength{\belowdisplayskip}{3pt}
\setlength{\abovedisplayshortskip}{2pt}
\setlength{\belowdisplayshortskip}{2pt}
\begin{equation}
\tilde d_{i,t} = d_{i,t} + \Delta d_{i,t}
\label{eq:tilde_d}
\end{equation}
\begin{equation}
\tilde p^{\mathrm{PV}}_{i,t} = p^{\mathrm{PV}}_{i,t} + \Delta p^{\mathrm{PV}}_{i,t}
\label{eq:tilde_pv}
\end{equation}
\endgroup
To maintain stealth, Eqs.~\eqref{eq:uncert_bounds_d} and \eqref{eq:uncert_bounds_pv} bound the perturbations within the forecast uncertainty margins, avoiding implausibly large deviations, while Eqs.~\eqref{eq:net_zero_d} and \eqref{eq:net_zero_pv} preserve the system-wide demand and PV totals at each time step, preventing detectable aggregate mismatches.
\begingroup
\setlength{\abovedisplayskip}{3pt}
\setlength{\belowdisplayskip}{3pt}
\setlength{\abovedisplayshortskip}{2pt}
\setlength{\belowdisplayshortskip}{2pt}
\begin{equation}
-\bar\Delta d_{i,t} \le \Delta d_{i,t} \le \bar\Delta d_{i,t}
\label{eq:uncert_bounds_d}
\end{equation}
\begin{equation}
-\bar\Delta p^{\mathrm{PV}}_{i,t} \le \Delta p^{\mathrm{PV}}_{i,t} \le \bar\Delta p^{\mathrm{PV}}_{i,t}
\label{eq:uncert_bounds_pv}
\end{equation}
\begin{equation}
\sum_{i\in\mathcal{I}} \Delta d_{i,t}=0
\label{eq:net_zero_d}
\end{equation}
\begin{equation}
\sum_{i\in\mathcal{I}} \Delta p^{\mathrm{PV}}_{i,t}=0
\label{eq:net_zero_pv}
\end{equation}
\endgroup
Together, Eqs.~\eqref{eq:upper_obj}--\eqref{eq:net_zero_pv} define the attacker's objective and feasible manipulation set.
 \subsection{Market Structure}\label{seclower}
Given the manipulated demand and PV inputs from the upper level, the market operator re-clears the interconnected multi-community markets by minimizing the total operating cost:
\begin{equation}\label{eq:lower_obj}
\begin{split}
\min \sum_{t\in\mathcal{T}} \Big[
\sum_{i\in\mathcal{I}} b^{\mathrm{DG}}_{i} P^{\mathrm{DG}}_{i,t}
+\sum_{i\in\mathcal{I}} \big(\pi^{\mathrm{grid}}_{t} P^{\mathrm{grid}}_{i,t}
-\pi^{\mathrm{FIT}}_{t} S^{\mathrm{grid}}_{i,t}\big) \\
{}+\sum_{c\in\mathcal{C}}\sum_{i\in\mathcal{I}_c}
\lambda_c^{\mathrm{com}} \big(P^{\mathrm{com}}_{i,t}+S^{\mathrm{com}}_{i,t}\big)
+\lambda^{\mathrm{int}} \sum_{c\in\mathcal{C}}
\sum_{\substack{c'\in\mathcal{C}\\ c'\neq c}} T_{c,c',t}
\Big]
\end{split}
\end{equation}
Here, \mbox{$P_{i,t}^{DG}$}, \mbox{$P_{i,t}^{grid}$},
\mbox{$S_{i,t}^{grid}$}, \mbox{$P_{i,t}^{com}$},
\mbox{$S_{i,t}^{com}$}, and \mbox{$T_{c,c',t}$} denote DG output,
grid purchase/sale, community purchase/sale, and inter-community trade,
respectively. The objective minimizes DG, net grid, community, and
inter-community trading costs, where \mbox{$b_i^{DG}$} is the DG
marginal cost, and \mbox{$\pi_t^{grid}$}, \mbox{$\pi_t^{FIT}$},
\mbox{$\lambda_c^{com}$}, and \mbox{$\lambda^{int}$} denote the grid
purchase price, feed-in tariff, community trading fee, and
inter-community trading fee, respectively. Since
\mbox{$\pi_t^{grid}>\pi_t^{FIT}$}, simultaneous grid purchase and sale
is not cost-optimal; thus, \mbox{$P_{i,t}^{grid}$} and
\mbox{$S_{i,t}^{grid}$} are mutually exclusive at the optimum.

The power balance at each bus is enforced by Eq. (\ref{eq:power_balance}), where the manipulated demand is supplied by a combination of dispatchable generation, manipulated PV generation, grid imports, and community-market purchases, while any surplus can be allocated to community-market sales or grid exports.
\begin{equation}\label{eq:power_balance}
\tilde{d}_{i,t} + S^{\mathrm{com}}_{i,t} + S^{\mathrm{grid}}_{i,t}
=
P^{\mathrm{com}}_{i,t} + P^{\mathrm{grid}}_{i,t} + P^{\mathrm{DG}}_{i,t} + \tilde{p}^{\mathrm{PV}}_{i,t}
\end{equation}

Equations~\eqref{eq:p2p_balance}--\eqref{eq:dg_q} define the community-market balance and generator operating limits. Specifically, Eq. \eqref{eq:p2p_balance} balances the aggregated net community-market purchases and sales of the players in each community through the directed inter-community trading, while Eqs. \eqref{eq:dg_p} and \eqref{eq:dg_q} bound the active and reactive power outputs of dispatchable generators.
\begingroup
\setlength{\abovedisplayskip}{3pt}
\setlength{\belowdisplayskip}{3pt}
\setlength{\abovedisplayshortskip}{2pt}
\setlength{\belowdisplayshortskip}{2pt}
\begin{equation}\label{eq:p2p_balance}
\sum_{i\in\mathcal{I}_c}\left(P^{\mathrm{com}}_{i,t}-S^{\mathrm{com}}_{i,t}\right)
=
\sum_{\substack{c'\in\mathcal{C}\\ c'\neq c}} T_{c',c,t}
-
\sum_{\substack{c'\in\mathcal{C}\\ c'\neq c}} T_{c,c',t}
\end{equation}
\begin{equation}\label{eq:dg_p}
\underline{P}^{\mathrm{DG}}_{i} \le P^{\mathrm{DG}}_{i,t} \le \overline{P}^{\mathrm{DG}}_{i}
\end{equation}
\begin{equation}\label{eq:dg_q}
\underline{Q}^{\mathrm{DG}}_{i} \le Q^{\mathrm{DG}}_{i,t} \le \overline{Q}^{\mathrm{DG}}_{i}
\end{equation}
\endgroup
The distribution-network constraints in Eqs.~\eqref{eq:pi}--\eqref{eq:thetalim} follow the LPF-D formulation in \cite{opf}, where \(G_{ij}\) and \(B_{ij}\) are the conductance and susceptance of branch \((i,j)\in\mathcal{E}\). These constraints represent linearized nodal injections, network power balance, and operating limits.
\begingroup
\setlength{\abovedisplayskip}{2pt}
\setlength{\belowdisplayskip}{2pt}
\setlength{\abovedisplayshortskip}{2pt}
\setlength{\belowdisplayshortskip}{2pt}
\begin{equation}\label{eq:pi}
P_{i,t} =
\sum_{\substack{j\in\mathcal{I}\\ j\neq i}}
\Big[
B_{ij}\big(\theta_{i,t}-\theta_{j,t}\big)
+G_{ij}\big(V_{i,t}-V_{j,t}\big)
\Big]
\end{equation}
\begin{equation}\label{eq:qi}
Q_{i,t} =
\sum_{\substack{j\in\mathcal{I}\\ j\neq i}}
\Big[
-G_{ij}\big(\theta_{i,t}-\theta_{j,t}\big)
+B_{ij}\big(V_{i,t}-V_{j,t}\big)
\Big]
\end{equation}
\begin{equation}\label{eq:pnode}
P_{i,t} =
\frac{P^{\mathrm{DG}}_{i,t}+\tilde p^{\mathrm{PV}}_{i,t}-\tilde d_{i,t}}{S^{\mathrm{base}}},
\qquad i \neq i^{\mathrm{ref}}
\end{equation}
\begin{equation}\label{eq:qnode}
Q_{i,t} =
\frac{Q^{\mathrm{DG}}_{i,t}-\tan(\varphi_i)\,\tilde d_{i,t}}{S^{\mathrm{base}}},
\qquad i \neq i^{\mathrm{ref}}
\end{equation}
\begin{equation}\label{eq:ptotal}
\sum_{i\in\mathcal{I}} P_{i,t}=0
\end{equation}
\begin{equation}\label{eq:qtotal}
\sum_{i\in\mathcal{I}} Q_{i,t}=0
\end{equation}
\begin{equation}\label{eq:vlim}
\underline{V}\le V_{i,t}\le \overline{V}
\end{equation}
\begin{equation}\label{eq:thetalim}
\underline{\theta}\le \theta_{i,t}\le \overline{\theta}
\end{equation} 
\endgroup
Although LPF-D approximates AC power flow, its convexity ensures Karush--Kuhn--Tucker (KKT)
sufficiency while preserving the voltage sensitivity relevant to FDI impacts.
Together, Eqs.~(\ref{eq:lower_obj})--(\ref{eq:thetalim}) define the lower-level problem,
whose voltage profile $\mathbf{V}$ is used in Eq.~(\ref{eq:upper_obj}).
\begin{equation}
\left\{
\begin{aligned}
\max \quad 
& \sum_{t\in\mathcal{T}}\sum_{i\in\mathcal{I}} u^V_{i,t} \\
\text{s.t.}\quad 
& \sum_{t\in\mathcal{T}}\sum_{i\in\mathcal{I}}
\left(u^d_{i,t}+u^{PV}_{i,t}\right)\le \varepsilon, \\
& u^V_{i,t}\ge V_{i,t}-V^{\mathrm{ref}}, \quad
u^V_{i,t}\ge V^{\mathrm{ref}}-V_{i,t}, \\
& u^V_{i,t}\le (V_{i,t}-V^{\mathrm{ref}})+M_V(1-z^V_{i,t}), \\
& u^V_{i,t}\le (V^{\mathrm{ref}}-V_{i,t})+M_V z^V_{i,t}, \\
& u^d_{i,t}\ge \pm \Delta d_{i,t}, \qquad
u^{PV}_{i,t}\ge \pm \Delta p^{PV}_{i,t}, \\
& z^V_{i,t}\in\{0,1\}, \\
& \text{upper-level constraints}, \\
& \text{KKT conditions of the lower-level problem.}
\end{aligned}
\right.
\label{eq:singlelevel_eps}
\end{equation}
\subsection{Solution Methodology}\label{sec:solution}
The linear lower-level problem is replaced by its Karush--Kuhn--Tucker
(KKT) conditions, with complementarity linearized using big-$M$
constraints and binary variables, yielding a mixed-integer linear program
(MILP). The big-$M$ values use physical, market-price, and dual-variable
bounds, with time-dependent values where applicable; auxiliary variables
linearize the absolute-value terms. The conflicting upper-level
objectives are handled by the $\varepsilon$-constraint method~\cite{eps},
maximizing cumulative voltage deviation subject to a detectability
bound. The resulting single-level MILP inz
Eq.~(\ref{eq:singlelevel_eps}) is implemented in Pyomo and solved with
Gurobi; varying $\varepsilon$ traces the impact--detectability Pareto
frontier.\section{Results and Discussion}\label{sec:results}
\subsection{System Setup and Data Description}
\begin{figure}[!b]
\centering
\includegraphics[width=\columnwidth]{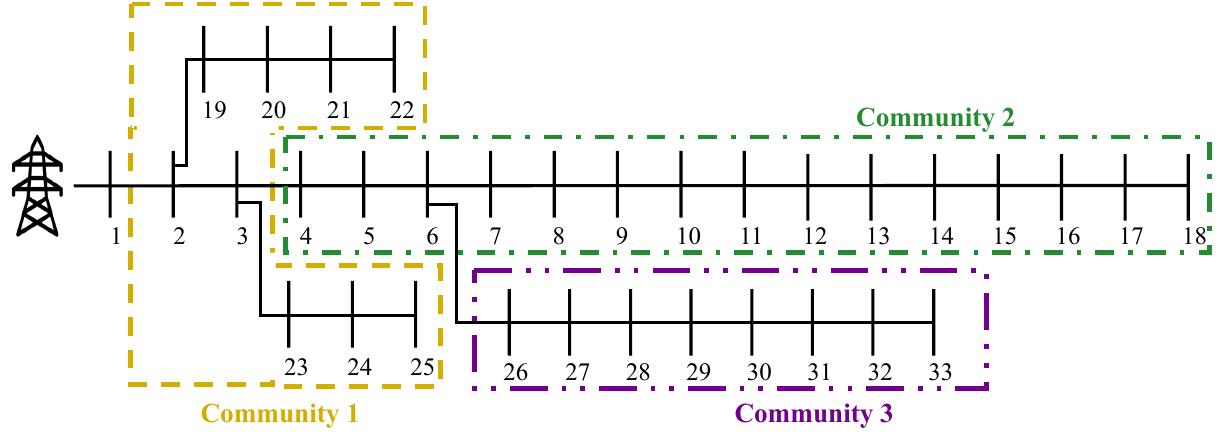}
\caption{IEEE 33-bus distribution system with three autonomous communities}
\label{fig:ieee33}
\end{figure}

\begin{figure}[!t]
\centering
\includegraphics[width=\columnwidth]{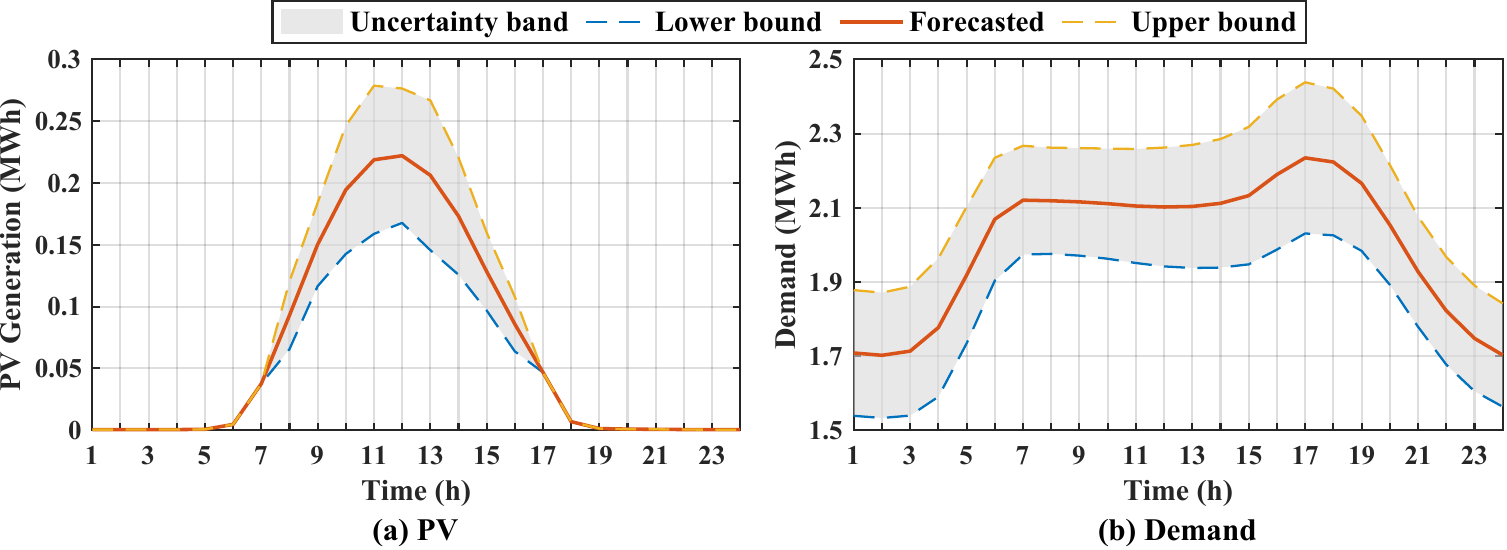}
\caption{Forecasted profiles with uncertainty bounds defining the admissible attack space: (a) PV generation; (b) demand.}
\label{fig:pro}
\end{figure}

The proposed framework is evaluated on the IEEE 33-bus
distribution network shown in Fig.~\ref{fig:ieee33}, with
three PV units and six dispatchable DGs located at the buses
listed in Table~\ref{tab:dg_data}. Historical price and
demand data from Sweden's SE3 bidding zone are obtained
from~\cite{ent}, with demand scaled to the test-system size
and the feed-in tariff set to 50\% of the market price. PV
generation is derived from Stockholm irradiance data from
Solcast~\cite{solcast} using the PVWatts
method~\cite{dobos2014pvwatts}. An xLSTM forecasting model \cite{repo-demo} trained on these load and PV series yields average upper-quantile deviations of approximately 4\% for load and 24\% for PV. September~30,~2024 is selected as the representative peak warm-season case-study day. Fig.~\ref{fig:pro} shows the
forecasted load and PV profiles with 80\% prediction
intervals used to define the admissible attack space.
Although only one representative case-study day and one
community partition are shown, the framework is directly
applicable to other distribution systems and community
configurations.
\begin{table}[!b]
\centering
\caption{Characteristics of Distributed Energy Resources}
\label{tab:dg_data}
\scriptsize
\setlength{\tabcolsep}{8pt}
\renewcommand{\arraystretch}{1.0}
\begin{tabular}{cccccc}
\toprule[0.5mm]
\toprule[0.1mm]
\textbf{Unit} & \textbf{Type} & \textbf{Bus} & \textbf{Com.} & \textbf{Capacity (kW)} & $b^{\mathrm{DG}}$ (SEK/kWh) \\
\midrule
1 & PV    & $i_{3}$  & $c_{1}$ & 400 & -- \\
2 & PV    & $i_{13}$ & $c_{2}$ & 545 & -- \\
3 & PV    & $i_{29}$ & $c_{3}$ & 200 & -- \\
4 & Disp. & $i_{3}$  & $c_{1}$ & 700 & 0.16\\
5 & Disp. & $i_{9}$  & $c_{2}$ & 760 & 0.12\\
6 & Disp. & $i_{17}$ & $c_{2}$ & 600& 0.13\\
7 & Disp. & $i_{21}$ & $c_{1}$ & 550 & 0.14 \\
8 & Disp. & $i_{29}$ & $c_{3}$ & 450 & 0.11 \\
9 & Disp. & $i_{32}$ & $c_{3}$ & 560 & 0.10\\
\bottomrule[0.1mm]
\bottomrule[0.5mm]
\end{tabular}
\end{table}

\subsection{Network Analysis Under Designed Attack}
\begin{figure}[!t]
    \centering
    \includegraphics[width=\columnwidth]{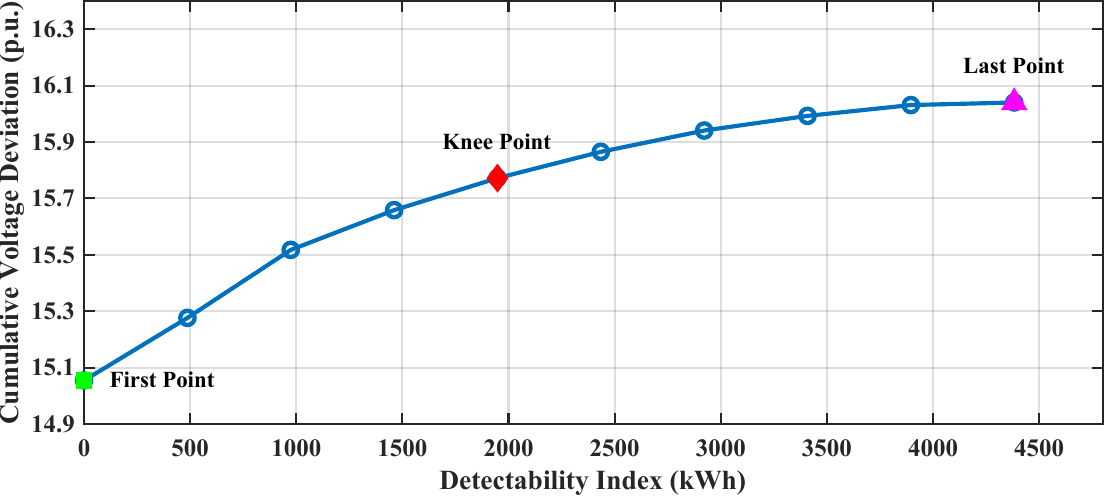}
    \caption{Trade-off between cumulative voltage deviation and detectability index along the Pareto frontier.}
    \label{fig:pareto}
\end{figure}

\begin{figure}[!t]
    \centering
    \includegraphics[width=\columnwidth]{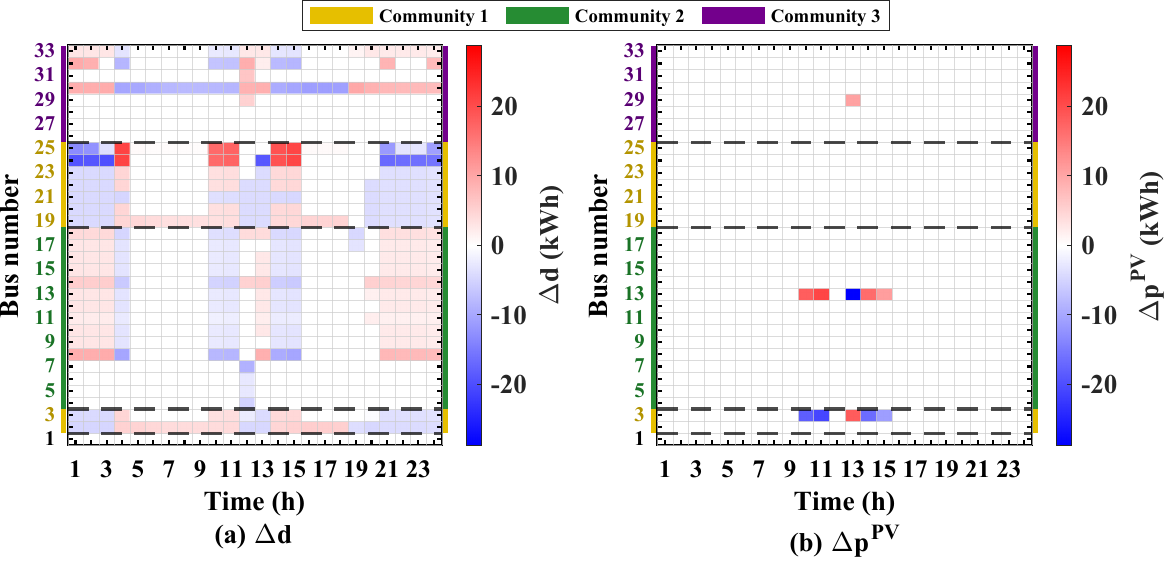}
    \caption{Spatiotemporal pattern of the optimized FDI: (a) demand perturbation $\Delta d_{i,t}$ and (b) PV perturbation $\Delta p^{PV}_{i,t}$.}
    \label{fig:man}
\end{figure}

\begin{figure}[!t]
    \centering
    \includegraphics[width=\columnwidth]{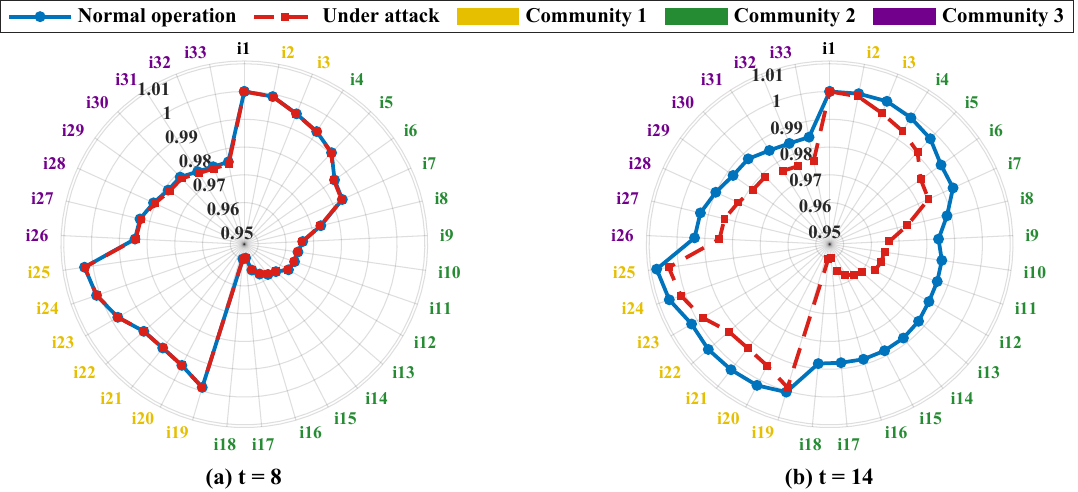}
    \caption{Voltage profiles under normal operation and attack: (a) $t=8$ and (b) $t=14$.}
    \label{fig:voltprof}
\end{figure}

This section presents the trade-off between detectability and
physical impact, together with the spatiotemporal placement of
the optimized FDI. Fig.~\ref{fig:pareto} shows the Pareto
frontier obtained by varying $\varepsilon$ in
Eq.~(\ref{eq:singlelevel_eps}), where detectability is measured
by the aggregate injected perturbations (kWh) and physical
impact by cumulative voltage deviation (p.u.). All points on the
frontier are valid Pareto-optimal solutions. The knee point,
identified as the solution with the maximum perpendicular
distance from the line connecting the minimum-detectability
and maximum-impact endpoints, captures 72.8\% of the
achievable voltage-deviation increase (from 15.05 to
15.77~p.u., a 4.77\% rise) while requiring only 44.4\% of the
maximum detectability, whereas the maximum-impact solution
raises the deviation to 16.04~p.u. (6.55\%) at full
detectability. The knee point offers the most efficient
trade-off for the attacker because it captures most of the
achievable voltage-security degradation at less than half of
the maximum detectability, while for the operator it marks a
critical regime in which moderately detectable attacks already
produce substantial physical impact. The knee point is
therefore adopted for the subsequent analysis.
Fig.~\ref{fig:man} shows that the optimized FDI is
concentrated in specific buses and time periods rather than
uniformly distributed. Demand and PV perturbations appear
mainly in parts of Communities~1 and~2, where higher demand,
larger installed PV capacities (400~kW and 545~kW,
respectively), and wider uncertainty bounds allow larger
admissible manipulations. However, only a subset of these
buses is targeted, indicating that the most influential attack
locations are not necessarily those with the highest load or
generation, but those where local injections have a stronger
effect on voltage. This sensitivity depends on network
topology and the prevailing operating condition. Accordingly,
the most effective bus--time locations do not necessarily
coincide with the peak PV and demand hours ($t=12$ and
$t=17$); indeed, the largest impact occurs at $t=14$. At
$t=14$, when the largest perturbations occur, the per-hour cumulative voltage deviation increases by 20.8\% (from 0.2767 to
0.3343), whereas at $t=8$, when the perturbations are
minimal, the increase is only 0.93\% (from 0.6867 to 0.6931)
(see Fig.~\ref{fig:voltprof}). This uneven temporal impact arises because voltage
sensitivity to a given perturbation varies with operating
conditions, network loading, and local resource dispatch.
Consequently, under a fixed detectability budget,
voltage-security degradation depends more on the
spatiotemporal placement of perturbations than on
distributing them uniformly across buses and time periods.

\subsection{Community-based Market Performance Assessment}
As shown in Table~\ref{tab:community_compare}, the coordinated FDI attack affects the three communities unevenly despite the system-level zero-sum constraint at each time slot. Within each community, locally matched purchases and
sales constitute intra-community trade, while any remaining
surplus or deficit is settled through inter-community exchange
or, if needed, grid imports or exports. Because the falsified
forecasts are distributed unevenly across communities, the
resulting market re-clearing produces asymmetric changes in
trading patterns, grid dependence, and operating cost.

Compared with normal operation, Community~1 remains
net-importing under attack, with part of its deficit shifted
from the main grid to neighboring communities. Community~2
experiences the largest market shift, with increased
inter-community exports, higher grid imports, and a 1.61\%
increase in operating cost. By contrast, Community~3 shows
only limited adjustment, with a small rise in inter-community
exports and a slight reduction in operating cost. Across all
three communities, DG dispatch changes by less than 0.25\%,
indicating that the attack mainly reshapes market exchanges
and grid dependence rather than total local generation.

The stronger economic impact on Community~2 stems from
its role as the marginal adjustment channel in the attacked
interconnected multi-community market re-clearing. The falsified forecasts
weaken internal supply--demand matching within
Community~2, so a larger residual imbalance must be
resolved through external exchanges. Because Community~1
remains net-importing and Community~3 already provides the
dominant share of inter-community exports under normal
operation, most of the incremental adjustment is absorbed by
Community~2, whose inter-community exports rise from
941~kWh to 1{,}227~kWh (30.4\%). Since local generation
changes only marginally, the additional energy needed to
support both Community~2's own balance and its enlarged
export obligation is drawn from the main grid, explaining the
concurrent rise in Community~2's grid imports from
6{,}434~kWh to 6{,}755~kWh. Consistent with Table~I, this
asymmetry reflects the underlying resource layout:
Community~1's higher DG marginal costs and weaker local
supply position leave it dependent on imports, Community~3's
lower marginal costs and stronger surplus position make it
the dominant exporter, and Community~2's intermediate DG
marginal costs together with the largest flexible DG and PV
capacity (1{,}905~kW) make it the natural absorber of the
marginal adjustment. As a result, although the attacker
optimizes only for cumulative voltage deviation, the induced
market re-clearing concentrates the main economic burden on
Community~2.
\begin{table}[!t]
\centering
\caption{Community-level market outcomes under normal operation and under attack.}
\label{tab:community_compare}
\setlength{\tabcolsep}{2.5pt}
\renewcommand{\arraystretch}{1.05}
\resizebox{\columnwidth}{!}{
\begin{tabular}{cccccccc}
\toprule[0.5mm]
\toprule[0.1mm]
\textbf{Case} & \textbf{Com.} & \makecell{\textbf{DG}\\\textbf{(kWh)}} & \makecell{\textbf{Intra-com.}\\\textbf{trade (kWh)}} & \makecell{\textbf{Inter-com.}\\\textbf{trade (kWh)}} & \makecell{\textbf{Grid imp.}\\\textbf{(kWh)}} & \makecell{\textbf{Grid exp.}\\\textbf{(kWh)}} & \makecell{\textbf{Cost}\\\textbf{(SEK)}} \\
\midrule
\multirow{3}{*}{Normal}
 & C1 & 10{,}299 & 6{,}044 & 4{,}216\textsuperscript{i} & 7{,}433 & 3{,}382 & 1{,}383 \\
 & C2 & 15{,}810 & 9{,}180 & 941\textsuperscript{e} & 6{,}434 & 5{,}066 & 1{,}370 \\
 & C3 & 14{,}854 & 4{,}992 & 3{,}275\textsuperscript{e} & 3{,}979 & 3{,}951 & 1{,}008 \\
\midrule
\multirow{3}{*}{Attack}
 & C1 & 10{,}275 & 6{,}093 & 4{,}554\textsuperscript{i} & 7{,}001 & 3{,}304 & 1{,}379 \\
 & C2 & 15{,}821 & 8{,}977 & 1{,}227\textsuperscript{e} & 6{,}755 & 5{,}066 & 1{,}392 \\
 & C3 & 14{,}844 & 4{,}887 & 3{,}327\textsuperscript{e} & 4{,}116 & 4{,}029 & 995 \\
\bottomrule[0.1mm]
\bottomrule[0.5mm]
\multicolumn{8}{l}{\scriptsize\textsuperscript{i}\,import;\; \textsuperscript{e}\,export.}
\end{tabular}}
\end{table}

\section{Conclusion}\label{sec:conclusion}
This paper proposes a bilevel optimization framework to assess coordinated FDI attacks on demand and PV forecasts in interconnected multi-community electricity markets. Case studies on the IEEE 33-bus system reveal a trade-off between detectability and physical impact, showing that bounded, stealth-constrained manipulations can reduce voltage-security margins. Vulnerability depends strongly on the spatiotemporal placement of perturbations rather than their uniform distribution across buses and time. Despite system-level zero-net FDI, the attack produces uneven community-level effects by redistributing inter-community exchanges, grid reliance, and economic outcomes. These findings motivate detection methods targeting spatiotemporally concentrated anomalies rather than aggregate consistency alone. Future work will consider robust market clearing under adversarial perturbations and detection schemes exploiting the spatiotemporal sparsity of optimal attacks.
\bibliographystyle{IEEEtran}
\bibliography{ref}
\end{document}